\documentclass[11pt, draftclsnofoot, onecolumn]{IEEEtran}
\usepackage{url}
\usepackage{xcolor}
\usepackage{cite,graphicx,amsmath,amssymb}
\usepackage{subfigure}
\usepackage{multicol}

\usepackage{fancyhdr}
\usepackage{mdwmath}
\usepackage{mdwtab}
\usepackage{amsthm}
\usepackage{multirow}

\usepackage{booktabs} % For professional table
\usepackage{multirow} % For professional table
\usepackage{tabularx} % For table to span the column width
\newcolumntype{Y}{>{\centering\arraybackslash}X} % For table to span the column width

\usepackage[ruled,vlined]{algorithm2e}

\newtheorem{theorem}{Theorem}

\newtheorem{lemma}{Lemma}

\newtheorem{corollary}{Corollary}

\allowdisplaybreaks[4]
\usepackage{hyperref}
\hypersetup{
	colorlinks=true,
    allcolors = black
}

\makeatletter
\def\ScaleIfNeeded{%
\ifdim\Gin@nat@width>\linewidth \linewidth \else \Gin@nat@width
\fi } \makeatother

\definecolor{coolgrey}{rgb}{0.55, 0.57, 0.67}
\definecolor{airforceblue}{rgb}{0.36, 0.54, 0.66}
\definecolor{americanrose}{rgb}{1.0, 0.01, 0.24}
\definecolor{amber}{rgb}{1.0, 0.49, 0.0}

\begin{document}

\title{ Minimum-Delay Routing for Integrated Aeronautical \emph{Ad Hoc} Networks Relying on Passenger-Planes in the North-Atlantic Region}

\vspace{-3.0em}
\author{%{\color{coolgrey}
Jingjing Cui,~ %~\IEEEmembership{Member, IEEE},
Dong Liu,~%~\IEEEmembership{Member, IEEE},
Jiankang Zhang,~%~\IEEEmembership{Member, IEEE},
Halil Yetgin, ~%~\IEEEmembership{Member, IEEE},
%arumugam~nallanathan~and~%\ieeemembership{fellow,~ieee}
%and 
Soon Xin Ng ~%~\IEEEmembership{Senior Member} and
and Lajos Hanzo%~\IEEEmembership{Fellow,~IEEE}
%}
%\thanks{this work would like to acknowledge the financial support of the engineering and physical sciences research council projects ep/noo4558/1, ep/po34284/1, coalesce, of the royal society's global challenges research fund grant as well as of the european research council's advanced fellow grant quantcom.}
\vspace{-3.0em}
}

\maketitle
%\vspace{-10.5em}
%{\color{coolgrey}
\begin{abstract}

Relying on multi-hop communication techniques, aeronautical \emph{ad hoc} networks (AANETs) seamlessly  integrate ground base stations (BSs) and satellites into aircraft communications for enhancing the on-demand connectivity of planes in the air.  
In this  integrated AANET context  we investigate the shortest-path routing problem with the objective of  minimizing  the total delay  of the in-flight connection from the ground BS subject to certain minimum-rate constraints for all selected links in support of   low-latency and high-speed services.  Inspired by the best-first search and priority queue concepts, we  model the  problem formulated by a weighted digraph and find  the optimal route based on  the shortest-path algorithm. 
Our simulation results demonstrate that   aircraft-aided multi-hop communications are capable of  reducing the total delay of  satellite  communications,  when relying on  real historical flight data.

\end{abstract}
%\vspace{-0.5em}
%\begin{IEEEkeywords}
%Aeronautical \emph{ad hoc} networks (AANETs), shortest-path routing,   flight data driven results.
%\end{IEEEkeywords}

\vspace{-0.9em}
\section{Introduction}
\label{sec:Introduction}
\vspace{-0.5em}
With the proliferation of Internet services and applications,  it is desirable to provide high-speed broadband access  during  flights above the clouds.    
  As mentioned in Airbus' Global Market Forecast (GMF) \cite{GlobalForecast19},  passenger traffic growth would increase by about 50\% by the year 2038 and air traffic as a whole will grow at 4.3\% annually over the next 20 years.  This forecast   further inspires the wide roll-out  of in-flight Internet access in aeronautical systems.  However, since aircraft fly at a high speed, it is challenging to  integrate conventional terrestrial communication solutions into aeronautical systems. 

Existing aircraft  connectivity solutions can be broadly categorized into two classes \cite{Vondra17ComMag,Zhang19Proc}: satellite to aircraft communication (S2AC) and direct aircraft to ground communication (DA2GC).  At the time of  writing in-flight connectivity of aircraft   mainly depends on satellites. In particular,  often geostationary Earth orbit (GEO) satellites are   used for providing on-board connectivity for aircraft  as a benefit of  their near-global coverage, supporting longer-lasting connections than DA2GC and  aircraft-to-aircraft communication (A2AC) without the need for  handovers between GEO satellites. For example, GoGo has  more than 1,000 aircraft that are equipped with   in-flight satellite connectivity, relying on the  Gogo 2Ku system,  for improving the  passengers' on-board Wi-Fi experience \cite{Gogo19sat}. 
However, the large coverage area of GEO satellites comes at the cost of high latency as well as substantial power loss. In particular, GEO satellites are at a distance of 35,768km, hence they suffer from a propagation delay of approximately 120 ms one-way delay  from the ground to the satellite.

To circumvent the shortcomings of S2AC, DA2GC provides an alternative for providing low-latency, high-rate transmissions to  aircraft \cite{Vondra17ComMag}.  The European Aviation Network (EAN) has been designed to deliver up to 75 Mbps per cell by combining an S-band satellite and  the 4G LTE mobile terrestrial network \cite{EAN15}. However, the coverage area of  DA2GC is limited, since the ground BSs can only be deployed on dry land, while about $2/3$ of the earth's surface  is covered by water. Furthermore, no  ground BSs are available in remote airspaces,  such as the polar regions, deserts, dense forests, etc. In this context, aeronautical \emph{ad hoc} networking constitutes  a promising technique of extending the  DA2GC by enabling aircraft in the sky to act as relays for aiding data transmission.   In  \cite{Rieth19TVT},  the effects  of both  large-scale fading and  small-scale fading on   DA2GC  channels  were demonstrated in the C-band. As a further advance, a resource management scheme designed for satellite based terrestrial networks  was investigated in \cite{Fu20tvt}. 

A key issue of routing in integrated AANETs is the link selection from the set of  D2AC, S2AC  and A2AC channels  across the routes, whilst  meeting the specific  application-oriented performance. {\color{black}In this paper, we concentrate on a special class of routing problems -- namely the shortest-path routing of integrated AANETs, which is synonymous with minimizing the total service delay subject to per-connection quality of service (QoS) constraints.  } 
The  shortest-path routing problem formulated can be solved over the associated weighted digraph by invoking the  shortest path  algorithm based on the idea of Dijkstra's algorithm  \cite{Dijkstra1959},  which is capable of delivering the optimal solution. We characterize our proposed algorithm by harnessing three real flight datasets based on   flights over the North-Atlantic oceanic area. In particular, our results reveal that  the aircraft can  be  connected to the  ground BS  through AANETs between London Heathrow (LHR)  Airport  and New York's John F Kennedy (JFK) Airport. Furthermore, both our simulations and real flight-data driven results demonstrate that  the A2AC links are capable of  extending the DA2GC coverage with the aid of  low-delay transmission.

%\pagebreak[3]
\vspace{-0.7em}
\section{System model}

\subsection{Preliminaries of Pareto-optimal solutions}
\begin{figure} [htp!]
\centering
\includegraphics[trim={0.1cm 0.1cm 0.1cm 0.1cm},clip, width = 0.450\linewidth]{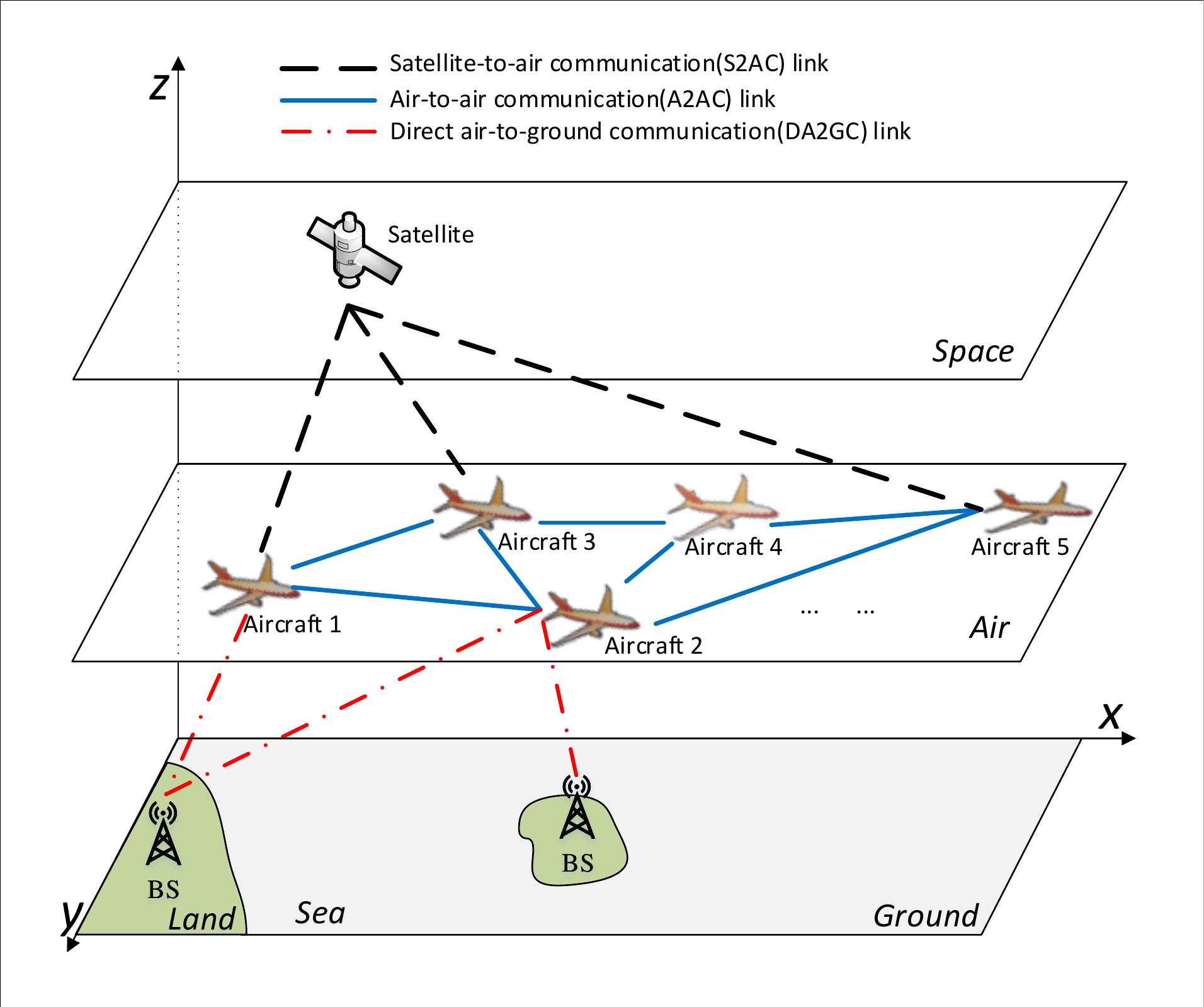}
  \vspace{-0.5em}  \caption{ \small Aircraft communication system by integrating ground and space communications. }
 \label{fig:aircraftsysmod}
   \vspace{-1.0em}
\end{figure}

We consider a ground-air-space integrated AANET comprising the ground layer, the aerial layer and the space layer  of  Fig.~\ref{fig:aircraftsysmod}, where aircraft can be connected to certain ground BSs either via direct communication or by multi-hop communication techniques. 
In particular, the aircraft can build communication links with other aircraft, ground BSs and satellites via A2AC, DA2GC and S2AC techniques, respectively,  in order to improve the on-board Internet experience of aircraft passengers.   We assume that the integrated  AANET considered is composed of $N_1$ aircraft, $N_2$ ground BS and $N_3$ satellites, where an example of the system model is illustrated in Fig. \ref{fig:aircraftsysmod}. Furthermore, given the mobility of aircraft during its flight its location changes rapidly.  For clarity, we denote the entities in the system encompassing aircraft, ground BSs and satellites as a set of nodes  $\mathcal{N} =  \mathcal{N}_1 \cup \mathcal{N}_2 \cup \mathcal{N}_3$ with $N = |\mathcal{N}|$, where $\mathcal{N}_1 = \{1,\cdots,N_1\}$, $\mathcal{N}_2 = \{1,\cdots,N_2\}$  and $\mathcal{N}_3 = \{1,\cdots,N_3\}$. 
%Similarly, we define $\mathcal{N}_a = \{1,\cdots,N_a\}$, $\mathcal{N}_g = \{1,\cdots,N_g\}$  and $\mathcal{N}_s = \{1,\cdots,N_s\}$ as the set of aircraft, ground BSs and satellites, respectively, such that $\mathcal{N} = \mathcal{N}_a \bigcup \mathcal{N}_g \bigcup \mathcal{N}_s$.   
%For convenience,  we assume that all aircraft use the same transmit power $P_{A}$ while  the ground BSs and satellites use the transmit power $P_{G}$ and $P_{S}$, respectively. Therefore, $P_i = P_{A}$ if $i \in \mathcal{N}_1$,  $P_i = P_{G}$ if $i \in \mathcal{N}_{2}$ and  $P_i = P_{S}$ if $i \in \mathcal{N}_2$. 
We  assume that the integrated AANET is operated in half-duplex mode as in \cite{Ernest19TWC} and the messages from different ground BSs are transmitted on different subchannels resulting in  no inter-link interference, representing an interference-free scenario.

 Since  aircraft typically fly $10$km above  the ground level,  they benefit from negligible scatterers and shadowing effects. Hence we assume that the communication links in the AANET have LoS propagation  \cite{Hofmann19ICC,Cuvelier18ISWCS,Vondra18Net}, where the path-loss between  node $i$ and node $j$, $i,j \in \mathcal{N}$,  can be expressed as
 {\setlength\abovedisplayskip{3pt} 
\setlength\belowdisplayskip{3pt}
{\small
\begin{align}
\label{eq:chamod}
h_{i,j} = \left( \frac{c}{4\pi d_{i,j}f_c} \right)^\alpha,
\end{align}}}
where $f_c$ is the carrier frequency  and $c = 3 \times 10^8$ m/s is the speed of light. Furthermore, $d_{i,j}$ is the distance between node $i$ as well as  node $j$ and $\alpha$ is the path-loss exponent.

\vspace{-0.7em}
\subsection{Delay Model}

In this paper,  the delay model relies on  the sum of the transmission and propagation delays of the individual connections. Moreover, we assume that the decode-and-forward (DF) relaying delay at each intermediate nodes is $D_{df} = 20$ ms.

Let us consider a communication link, where the message is sent from aircraft $i$ to aircraft $j$. Hence the   signal-to-noise ratio (SNR)  at the receiver of aircraft $j$ can be expressed as
{\setlength\abovedisplayskip{3pt} 
\setlength\belowdisplayskip{3pt}
{\small
\begin{align}
\phi_{i,j}  =\frac{P_i G_i^t G_j^r h_{i,j}}{\sigma^2}, \label{eq:snr}
\end{align}}}
where $\sigma^2$ denotes  the noise power. Furthermore,
$G_i^t$ and $G_j^r$ are the transmit antenna gain and the receive antenna gain,  respectively.
Therefore, the data rate in the link spanning from node $i$ to node $j$ can be expressed as
\begin{align}
C_{i,j} = B\log_2 \Big( 1 + \phi_{i,j}\Big),
\end{align}
where $B$ is the bandwidth allocated to the link.
Correspondingly, given the size $L$ of the data file to be transmitted in bits, the file-transfer delay between node $i$ and node $j$ can be calculated as
{\setlength\abovedisplayskip{3pt} 
\setlength\belowdisplayskip{3pt}
{\small
\begin{equation}
D_{i,j}^{tr} = \frac{L}{C_{i,j}}.
\label{eq:delaytrans}
\end{equation}}}

%\subsubsection{Propogation Delay}
The propagation delay is the time it takes for  the signal to travel at the speed of light through the communication link from a node to the  next one, which is given by 
{\setlength\abovedisplayskip{3pt} 
\setlength\belowdisplayskip{3pt}
{\small
\begin{equation}
D_{i,j}^{pr} = \frac{d_{i,j}}{c},
\label{eq:delayprop}
\end{equation}}}
where $D_{i,j}^{pr} = D_{j,i}^{pr}$. 
As a result, the total delay from node $i$ to node $j$ can be expressed as
{\setlength\abovedisplayskip{3pt} 
\setlength\belowdisplayskip{3pt}
{\small
\begin{equation}\label{eq:delaytot}
D_{i,j} = \begin{cases}  
	    D_{i,j}^{tr} + D_{i,j}^{pr}, & \text{if} ~j ~\text{is the target aircraft}, \\
	    D_{i,j}^{tr} + D_{i,j}^{pr} + D_{df}, & \text{otherwise} .
\end{cases}
\end{equation}}}

\vspace{-0.9em}
\section{Problem Formulation}
\label{sec:proform}
%{\color{coolgrey} Some insightful background explanations goes here.}

Due to the curvature of the Earth, the maximum  direct propagation  distance between two aircraft is limited. Hence in addition to the QoS constraints, the communication between two nodes is also restricted by  the radio-horizon,  denoted as $d_{vis}$, which is relying on the height of the two nodes.

Our objective is  to minimize the total  delay of the links selected for transmission by appropriately selecting the routes between the source BS  on the ground and the destination aircraft, which can be expressed as
{\setlength\abovedisplayskip{3pt} 
\setlength\belowdisplayskip{3pt}
\small
\begin{subequations}\label{eq:optprob}
\begin{align}
\min_{x_{i,j} \in \mathcal{X}}  \quad &  \sum_{i \in \mathcal{N}}\sum_{j\ne i,j \in \mathcal{N}}  D_{i,j} x_{i,j} \\
\mathrm{s.t.} \quad
 & x_{i,j}\phi_{i,j} \geq \phi_0,
								~i ~\text{is not the target aircraft},
 \label{eqc1:legitimatelinks}\\		
&d_{i,j} \le d_{vis}, \label{eqc2:visdis} 
 	\\
 \begin{split}
 &   \sum_{\substack{ j\ne i \\ j\in \mathcal{N}}}x_{i,j} - \sum_{\substack{j \ne i \\ j \in \mathcal{N}}}x_{j,i} = \begin{cases}
	1, & \text{if}~i  = s, \\
	-1, & \text{if}~i = d,\\
	0, &\text{otherwise},
	\end{cases}	
\end{split}	
	\label{eqc3:noloops}\\			
\begin{split}
& 		\sum_{j\ne i, j\in \mathcal{N}} x_{i,j}  \begin{cases}
	\leq 1, &\text{if} ~i \neq d, \\
	= 0, &\text{if}~i = d,
	\end{cases} 
	\end{split}\label{eqc4:nodedegrees} \\
	& \forall i,j \in \mathcal{N}, i\neq j ,
\end{align}
\end{subequations}}\noindent where $s$ and $d$ denotes the source and the destination, respectively.  $x_{i,j}$ is  the link indicator function used, where  we have  $x_{i,j} = 1$ if link $(i,j)$ is on the optimal route; Otherwise $x_{i,j} = 0$.
%$\mathcal{D} \subseteq \mathcal{N}_a$ denotes the set of the target aircraft that  the ground BSs want to communicate to and  
Furthermore, $\phi_0$ is a predefined SNR threshold to be exceeded for guaranteeing the received signal quality, while the constraints in  \eqref{eqc1:legitimatelinks} and \eqref{eqc2:visdis}  ensure that a communication link spanning from node $i$ to node $j$ exists.
Finally, constraints in  \eqref{eqc3:noloops} and \eqref{eqc4:nodedegrees}  ensure that the solution found from the problem formulated does indeed represent a legitimate path spanning from the ground BSs to the target aircraft.

\vspace{-0.7em}
\section{Solutions for Finding Optimal Routes}

As discussed in Section \ref{sec:proform}, the delay minimization problem considered in this paper is an optimal route finding problem between the source node and the destination, which can be transformed into the optimal route-finding problem on a weighted  digraph. Correspondingly, we first generate
 a weighted digraph based on the available network information. 
The initialization process is given in {\bf Algorithm \ref{alg:Init}}, in which  a weighted digraph $\mathcal{G}(\mathcal{V},\mathcal{E})$ is constructed based on the system information and  the constraints of \eqref{eqc1:legitimatelinks} as well as \eqref{eqc2:visdis}, where $\mathcal{V}$ and $\mathcal{E}$ represent the set of the nodes and the edges, respectively.  Note that the initialization process in {\bf Algorithm \ref{alg:Init}} is capable of reducing the size of the  weighted digraph generated by removing the redundant edges and nodes by considering the constraints in \eqref{eqc1:legitimatelinks} and the curvature of the earth in \eqref{eqc2:visdis}.

\begin{algorithm}
\small
\caption{Generated a weighted digraph based on AANETs}
\label{alg:Init}
\SetKwInOut{Input}{Input}\SetKwInOut{Output}{Output}
%\SetKwProg{Init}{Initialization}{}{}

\Input{The dataset for the number and the locations of aircraft, ground BSs and Satellites.
}
\Output{Generated weighted digraph of ANETs}
\SetKwInOut{Input}{Init.}
\Input{SNR threshold: $\phi_0$; Values of system parameters:  $G^t$,  $ G^r$,  $f_c$ and $B$\;
 % Values of system parameters: Transmit and receive antenna gain: $G^t$ and $ G^r$,  carrier friency: $f_c$ and bandwidth $B$. \\
}

%\tcp{Generate the weighted digraph $\mathcal{G}(\mathcal{V},\mathcal{E},\mathcal{W})$.}
	%Set $i = 0$\;
\Repeat{all nodes are visited}
{
%	$i = i + 1$\;
	For $i \in \mathcal{N}$,  calculate the received SNR at every other node $j$ from node $i$ using \eqref{eq:snr} denoted as $\boldsymbol{\phi}_{i} = \{\phi_{i,j}, ~ j \ne i ~\mathrm{and}~ j \in \mathcal{N}\}$\;
	\While{ $j \in \mathcal{N}$ and $j \ne i$}
	{	
	\If{$\phi_{i,j} \ge \phi_0$ and node $j$ is visible to node $i$ }{
		Calculate the delay based on \eqref{eq:delaytot} as the weight of the edge $e_{i,j}$\;
		Add edge $e_{i,j}$ and its weight $D_{i,j}$ into the graph $\mathcal{G}$\;}
	}
}
\end{algorithm} \vspace{-0.9em}
There are a number of approaches for finding the optimal route from the source to the destination such as Dijkstra's algorithm, dynamic programming as well as genetic algorithms. In this paper, we employ the best-first search strategy (also called priority-first search) of Dijkstra's algorithm for finding the optimal route of the problem in \eqref{eq:optprob}, which is summarized in {\bf Algorithm \ref{alg:Dijkstra}} .

Explicitly,  the ground BS and the target aircraft are defined to be the source and the destination node, respectively.  The total delay between any two nodes is treated as the `cost' of the link. In addition, the priority queue structure is used for storing the candidate nodes as well as the correlated cost information during the search process. 
The  procedure of  {\bf Algorithm \ref{alg:Dijkstra}}  starts by introducing a set $\psi = \{\psi_0,\cdots,\psi_N\}$ for storing the objective function (OF) values  from the source node $s$ to each of the other nodes during the search process. In $\psi$, all elements $\psi_v$, $v \in \mathcal{V}$, are initialized to $\infty$, except for $\psi_s = 0$.  Furthermore, in  {\bf Algorithm \ref{alg:Dijkstra}},  the priority queue $Q$ is used for storing the current leaves in the current search tree as well as their OF values in $\psi$, where the nodes in the priority queue $Q$ are sorted in an ascending order based on  the OF values. Specifically, the node with the minimum OF value has the top priority and thus will be first taken out.    Moreover, $\mathrm{prev[v]} = u$ indicates that the parent  of node $v$ is $u$ and thus the route spanning from $s$ to $d$ can be constructed  by visiting  $\mathrm{prev}$ recursively.
To speed up the search process over the solution space, we introduce an additional variable $\psi_{min}$ for storing the minimum delay spanning from the source node to the destination during the search process. Thus, $\psi_{min}$  provides an upper bound of the OF value, since problem \eqref{eq:optprob} is a minimization problem of the total delay. The step in line \ref{alg2:setmin} guarantees  that $\psi_{min}$ is always  the minimum OF value from $s$ to $d$.
Note that from the delay model of \eqref{eq:delaytot},  the total delay spanning from the source node to any other node is monotonically increasing with the number of nodes in the route. As a consequence, the leaf node $v$ in $Q$  associated with $\psi_v > \psi$ will be pruned  as seen  in line \ref{alg2:judgmin}, which is capable of significantly reducing the search space, especially for large networks.
% Instead of visiting  
% Instead of visiting all nodes in the digraph, the search process will stop once the destination is visited as shown in  line \ref{alg2:setmin} of Algorithm 2.  
However, in the worst-case scenario,   {\bf Algorithm \ref{alg:Dijkstra}} will have the same complexity as the standard Dijkstra algorithm \cite{Fredman87ACM}, which is of the order  of $\mathcal{O}(|\mathcal{E}| + |\mathcal{V}| \log |\mathcal{V}|)$.
 %the improved Dijkstra algorithm also implemented based on the priority queue structure.

\begin{algorithm}
\small
\caption{The shortest-path routing algorithm}
\label{alg:Dijkstra}
\SetKwInOut{Input}{Input}\SetKwInOut{Output}{Output}
%\SetKwProg{Init}{Initialization}{}{}

\Input{Weighted digraph of ANETs; The start node $s$ and the destination $d$\;}
\Output{The optimal route $R^*$ and the minimum delay $\psi^*$;}
%\tcp{Dijkstra's algorithm for finding the optimal route:}
%\tcc{Initialization of each node}
\SetKwInOut{Input}{Init.}
 \Input{ A priority queue $Q= \emptyset$\; }
 \nl Initialize $\psi= \{\psi_v,  \forall v \in \mathcal{V}\}$, where $\psi_v = 0$ if $v$ is $s$, and $\psi_v = \inf$ if otherwise. Set $\psi_{min}  = \inf$\;
 \nl Push $\psi_s$ and $s$ to $Q$\;

\nl \While{$Q$ is not empty}{

\nl {Pop the vertex $u$ and $\phi_u$\;} %\tcp*[f]{$\psi_u \le \psi_i,~ i \in Q$} \\ %with minimum distance\;
\nl \If{$u = d$} 
{
	$\phi_{min} = \min\{\phi_{min} ,\phi_u$\}\; \label{alg2:setmin}
}
\nl \While{$v$ is a neighbor of $u$}
{
  \nl  $\hat{\psi} = \psi_u + D[e_{u,v}]$\;
	\nl \If{$\psi_v >= \hat{\psi}$ and $\psi_v \le \psi_{min}$ \label{alg2:judgmin}}
	{
			\nl $\psi_v = \hat{\psi}$; 	$\mathrm{prev}[v] = u$\;
			\nl Store $\psi_{v}$ and $v$ into the $Q$\;
	}
	
}
}
\nl Constitute $R$ from $\mathrm{prev}$; $\psi^* = \psi_d$ and $R^*=R$\;
%\nl	 Go to the end of the algorithm\;\label{alg:pro1s2}
			
%\nl $\psi^* = \psi_d$ and $R^*=R$\;
\end{algorithm}\vspace{-0.9em}

\vspace{-0.7em}
\section{Simulations}
\label{sec:sim}

In this section, we evaluate the performance of the proposed algorithms by computer simulations. We first investigate the results based on simulations in Section  \ref{subsec:numresults}. Then we apply our algorithms to real flight datasets collected in the North-Atlantic region in Section \ref{subsec:dataresults} for validating the performance of the algorithms, which also reveals the potential benefits of AANETs in terms of improved connectivity.

\vspace{-0.5em}
\subsection{ Numerical Results}
\label{subsec:numresults}

To model flight paths and the satellite orbits above the Earth surface, the spherical  coordinate system denoted by $(r,\theta,\varphi)$ is considered, where the origin is located in the center of the earth. Furthermore, $r$ is the distance of a point from the origin,  while $\theta$  and $\varphi$ are the polar angle and the azimuthal angle of the point, respectively.
%Explicitly, $R$ is the distance of a point to the origin also called radial distance, $\theta$ is the polar angle which is the angle from the zenith direction to the point, and $\phi$ is the azimuthal angle measured from the azimuth reference direction to the orthogonal projection of the line segment to the point  on the reference plane.
As a result, the locations of the ground BS, aircraft and satellites can be denoted as $(r_{g},\theta_{g},\varphi_{g})$, $(r_{a},\theta_{a},\varphi_{a})$ and  $(r_{s},\theta_{s},\varphi_{s})$, respectively, where $r_{i} = H_{i} + r_{earth}$ for any $i \in \mathcal{N}$.
To characterise the proposed algorithm, we first consider the simple scenario of $N_g = N_s = 1$. Specifically,  $\theta_{g} = \varphi_{g} = 0$, which indicates that we treat the ground BS as the reference point of the system in the spherical coordinate system. The target aircraft is located at the point  $\theta_d = \frac{\pi}{4}$ and $\varphi_d = \frac{\pi}{6}$, while the GEO satellite  is located at the center of the region with  $\theta_{s} = \frac{\pi}{8}$ and  $\varphi_{s} = \frac{\pi}{12}$. Furthermore, the values of $\theta_a$ and $\varphi_a$ for the intermetiate aircraft  are randomly drawn from a uniform distribution with $\theta_{a} \in [0,\theta_d ] $ and  $\phi_{a} \in [0, \varphi_d]$, respectively. Hence,  all figures are plotted over 1000 realizations. Note that the distance between the ground BS and the target aircraft  is about 3300 km, and thus the ground BS cannot  communicate directly with the target aircraft due to the curvature of the Earth. Moreover, we assume that  the system operates at the mm-Wave frequency of $f_c = 31$ GHz and the noise power  is $\sigma^2 = -132$ dBm  \cite{Hofmann19ICC}.   The other parameters used in our simulations  are the same as those in  \cite{Vondra18Net}. Specifically,   the transmit powers of the ground BS,  of the aircraft and of  the satellite are  45dBm,  30dBm and 50dBm, respectively. Furthermore, we have $G_i^{t} = G_i^r = 25$dB for the ground BS and the aircraft, while   $G_i^{t} = G_i^r = 45$dB for the satellite along with $B = 200$MHz and $\phi_0 = 0$dB. The heights of the ground BS, the aircraft as well as the satellite are 50m,10.7 km and 35768 km, respectively. In particular, the 
 greatest  distance between two points $i$ and $j$ above  the horizon is calculated by  $d_{vis} < 3.57 (\sqrt{H_i} + \sqrt{H_j})$, where $d_{vis}$ is in kilometers and $H_i$ and $H_j$ are in meters.

\begin{figure} [t!]
\centering
\subfigure[Delay versus the number of  the intermediate aircraft.]{
\includegraphics[trim={0.3cm 0.1cm 0.1cm 0.1cm},clip,  width = 0.44\linewidth]{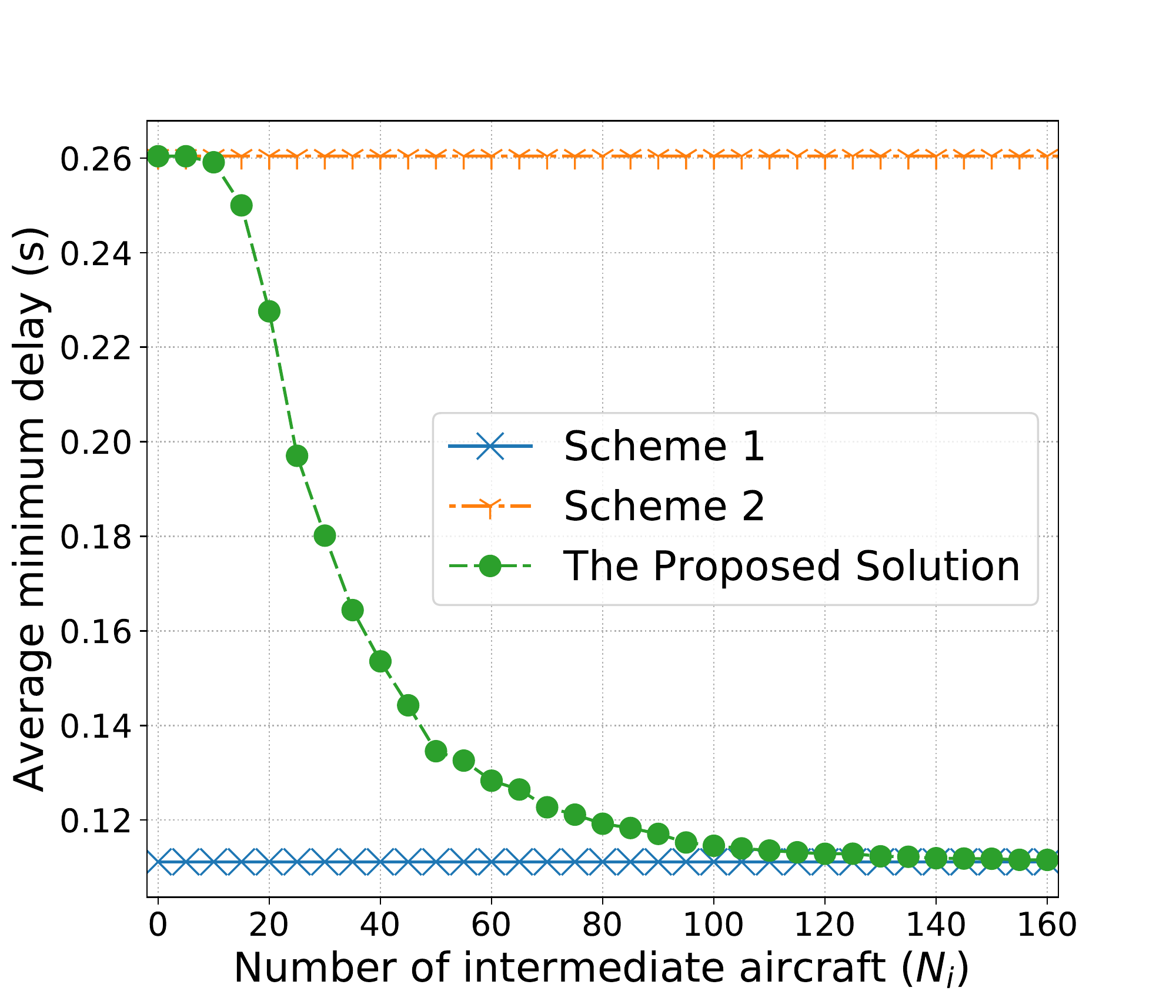}
   \label{fig:delayvsnodes}
   }
\subfigure[Hop counts versus the number of  the intermediate aircraft.]{
\includegraphics[trim={0.3cm 0.1cm 0.1cm 0.1cm},clip, width = 0.48\linewidth]{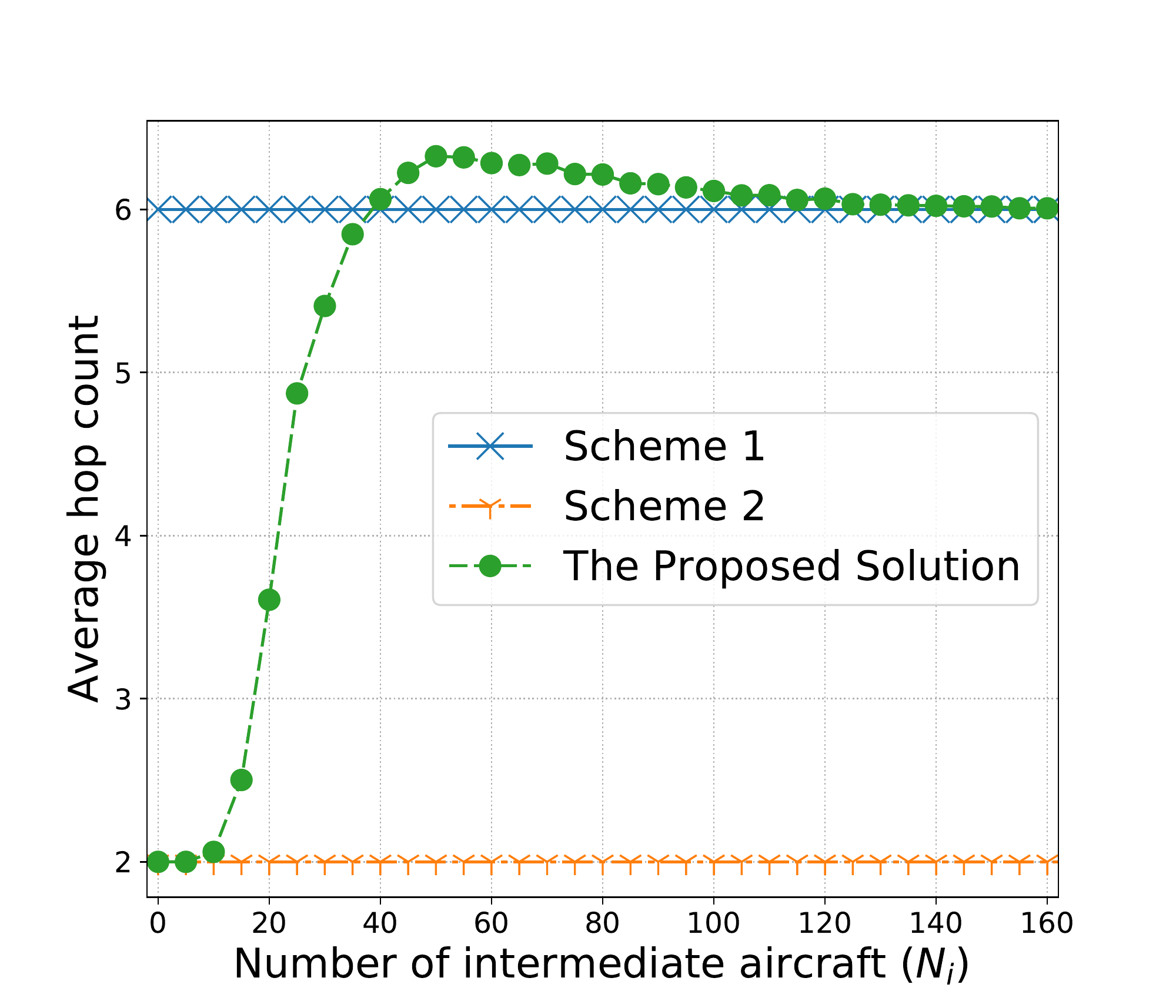}
\label{fig:hopsvsnodes}
}
  \vspace{-0.5em}  \caption{Comparisons of the total delay as well as the hop count for three different schemes.}
  \label{fig:delay_hopsvsnodes}
   \vspace{-1.0em}
    \vspace{-0.50em}
\end{figure}

\begin{figure} [t!]
\centering
\includegraphics[trim={0.3cm 0.1cm 0.1cm 0.1cm},clip, width = 0.5\linewidth]{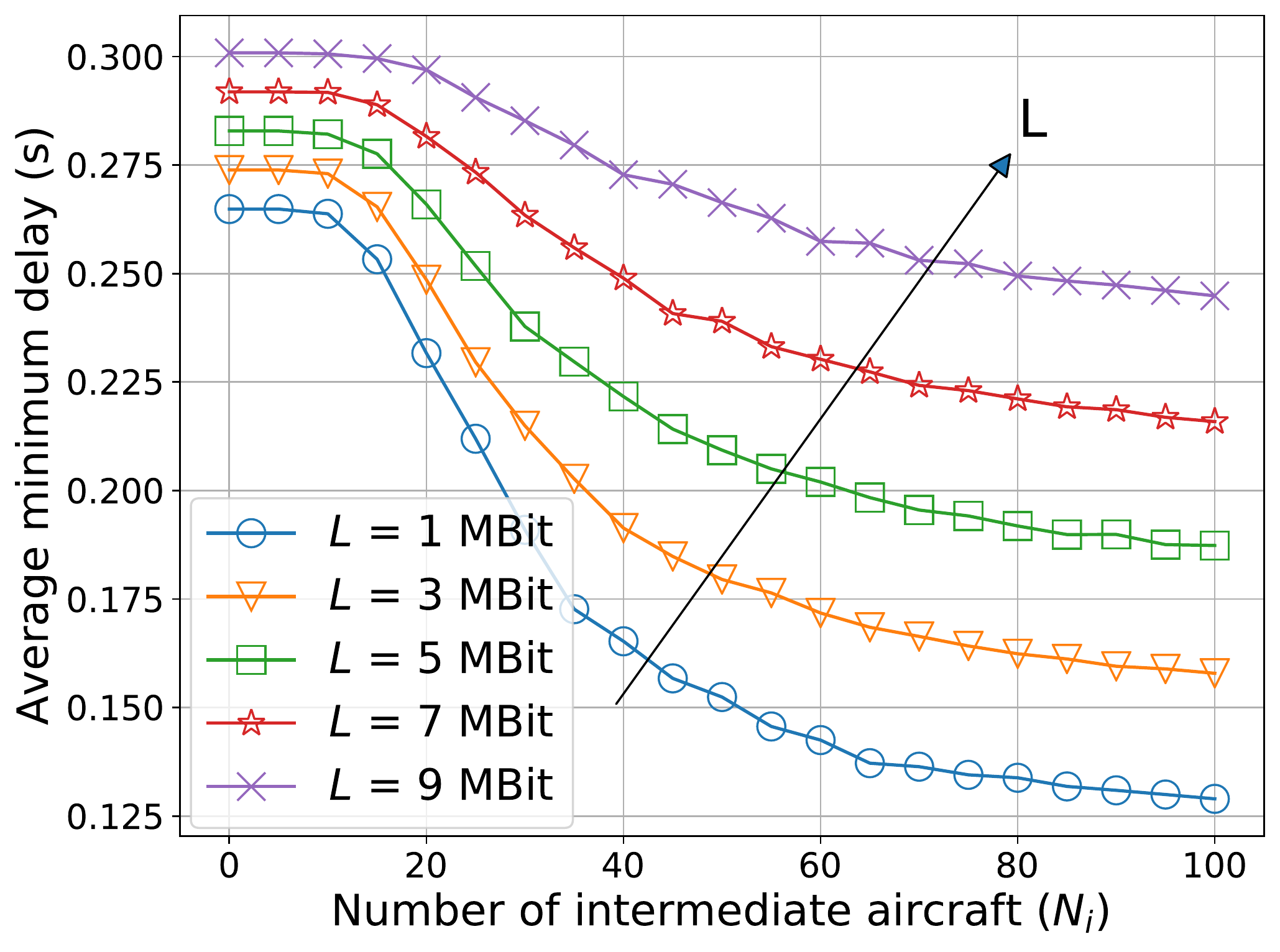}
 \vspace{-0.5em}  \caption{ \small{The impact of file size on the minimum delay.}}
  \label{fig:DvsPKL}
   \vspace{-1.0em}
   \vspace{-1.0em}
\end{figure}

Fig. \ref{fig:delay_hopsvsnodes} illustrates the total delay and the number of hops  in the shortest path\footnote{Note that  in this paper  the number of hops in a path denotes the number of edges nodes in the path. For instance, consider a route with four nodes like $s \rightarrow n_1 \rightarrow n_2 \rightarrow d$, the number of hops is 3.} versus the number of intermediate relaying aircraft with different schemes, where the transport block size of  $L = 9000$ bits is used as  in  \cite{3GPPR15}.  
For comparisons, we consider two different schemes. In Scheme 1,  we assume   that the ground BS is able to communicate with the target aircraft via A2AC links, which provides a lower bound of the end-to-end delay.  Note that the distance between the ground BS and the target aircraft is around 3300 km in the model considered, hence it requires  six hops at least due to the  constraint of the visible distance.  As a result, the lowest possible  delay  can be attained according to  \eqref{eq:delaytot}.  In  Scheme 2, we consider  satellite aided communications, which only has two hops due to the large coverage area of the GEO satellite.  Therefore, Fig. \ref{fig:delayvsnodes} illustrates how the total delay and the number of hops  in  the shortest path vary upon increasing the number $N_i$ of intermediate aircraft. 
 As seen from Fig. \ref{fig:delayvsnodes}, the total delay of the proposed solution is substantially reduced upon increasing the number of intermediate aircraft.  In particular, the minimum delay attained by the proposed algorithm  closely approaches  the lower bound of Scheme 1, when  $N_i \ge 120$.  This  indicates that aircraft aided multi-hop communications substantially  reduces the  end-to-end delay upon using  the proposed algorithm.    As  observed from  Fig. \ref{fig:hopsvsnodes},   the hop count of the proposed solution is lower than that of Scheme 1 when $N_i < 35$  owing to occasionally opting for the satellite link as part of the shortest path for avoiding any potential route breakage.  However, the hop count becomes higher than that of Scheme 1 when $N_i\le 40$. Furthermore,  the  hop count of the proposed solution  closely approaches that of  Scheme 1 when $N_i \le 120$ as in Fig. \ref{fig:delayvsnodes}, which implies that the hop count in the proposed solution  approaches the minimum number of  hops required, when the number of intermediate aircraft is high enough. 
Fig. \ref{fig:DvsPKL} illustrates the impact of the file size on the total delay.  As seen from Fig. \ref{fig:DvsPKL},   the total delay between the source on the ground and the target aircraft is reduced  upon increasing the number of aircraft, because this improves the probability of  selecting a lower-delay aircraft-aided multi-hop link instead of the two-hop, yet high-delay satellite link for reaching the target aircraft. 
%the total propagation delay is the sum of $11$ ms plus $80$ ms relaying delay yielding 91 ms.   
Therefore,  when there  is a shortage of   intermediate aircraft,  the ground BS communicates with the target  aircraft via the satellite, which results in a higher propagation delay compared to A2AC links.   In this context, the total propagation delay is dominated by that of  the satellite links. Hence the delay versus $N_i$ curve of Fig. \ref{fig:DvsPKL} remains relatively flat  when the number of aircraft is below about 20. 
Furthermore, 
we can observe that upon increasing  the  file size of $L$ from 1Mbit to 9Mbit, the delay difference between $N_i = 0$ and $N_i = 100$ is substantially increased. 
%This is mainly due to the  huge coverage, which results in  less hops required in the satellite aided transmission than the A2AC  aided transmission.   In consequence, there is a tradeoff between the  satellite aided transmission and the A2AC  aided transmission. 

Let us elaborate briefly on a simple scenario with all links having  an equal transmission rate of $C = 10$Mbps.  For the satellite aided transmission, the total uplink and downlink propagation delay of the satellite is about $D_{pr} = 240$ms, while the total uplink plus downlink  file-transfer delay  is  $D_{tr} = 40$ms for a $L=200$Kbit file. Finally, the DF relaying delay is assumed to be  $D_{df} = 20$ms. Thus, we can obtain the total delay of  $D^{S2AC} = 300$ms for an $L=100$Kbit file.   
By contrast,  for the A2AC  aided transmission relying on five hops in Scheme 1,  the total propagation delay is around $D_{pr} = 11$ ms for a distance of 3300km, the file-transfer  delay is $D_{tr} = 120$ms and  $D_{df} = 100$ms. Hence, we have the total delay of $D^{A2AC} = 231$ms.  In this case,  the A2AC  aided transmission outperforms  the satellite aided transmission. On the other hand, when the ground BS has an $L = 1$Mbit file for transmission, we have $D^{S2AC} = 460$ms while $D^{A2AC} = 711$ms, which results in a lower delay upon using the satellite than using the A2AC  link\footnote{Note that given a transmission rate $C$, there is a threshold $L_{th}$, where the A2AC aided transmission is beneficial if $L<L_{th}$, otherwise the satellite aided transmission has a lower delay. In particular, from the delay model of \eqref{eq:delaytot}, we have $L_{th} \approx \frac{3.658C}{100}$}.  
%Furthermore, the file-transfer of the satellite aided transmission is smaller than that of the A2AC aided transmission since the satellite has lager transmit power as well as the transmit and receive antenna gain than aircraft.
This reveals that aircraft aided multi-hop communications may have an  advantage in providing lower-delay  services  than satellite aided communications, provided that the file  is not excessively large. 

%%%%%%%%%%%%%%%%%%%%%%%%%%%%%%%%%%%%%%%%%%%%%%%
%Results on real flight data
\vspace{-0.7em}
\subsection{Flight Data Driven Results}
\label{subsec:dataresults}

In this subsection, we characterize the system performance for the  five busiest  TransAtlantic airlines using real historical flight data collected over the North-Atlantic region, which includes Delta Airline, United Airline, American Airline, British Airways and Lufthansa.
Specifically, these datasets contain the historical flight information of the area recorded at sampling intervals of  10s, where each entry of the flight contains the following information:  timestamp,  longitude,   latitude,  altitude and speed. 
The data in the first and second datasets are collected from  00:00 on 24 Dec. 2017 until 00:00 on 26 Dec.2017, which is typically the quietest day of the year. 
The first dataset, referred to  as Data-1,  contains the TransAtlantic flights between  LHR Airport and  JFK Airport  as in the previous example, which consists of 57 flights and 17,281 entries  for  each flight.  The second dataset, referred  as  Data-2, contains all TransAtlantic flights of the 5 busiest  TransAtlantic airlines, which consists of 381 flights and  17,281 entries  for  each flight. The data in the third dataset were collected   from  00:00 on 29 Jun. 2018 to 00:00  on 30 Jun. 2018, which is the busiest day of the year having the most flights.   Similar to the scenario of Data-2, the third dataset, Data-3, contains 649 flights and 8,641 entries for each flight.  Moreover, the file size used is $L = 200$Kbit and all the  other parameters used in our  simulations are the same as in Section \ref{subsec:numresults}.

\begin{figure}[!htp]
\centering
\subfigure[Results on Data-2 ]{
\includegraphics[ width = 0.45\linewidth]{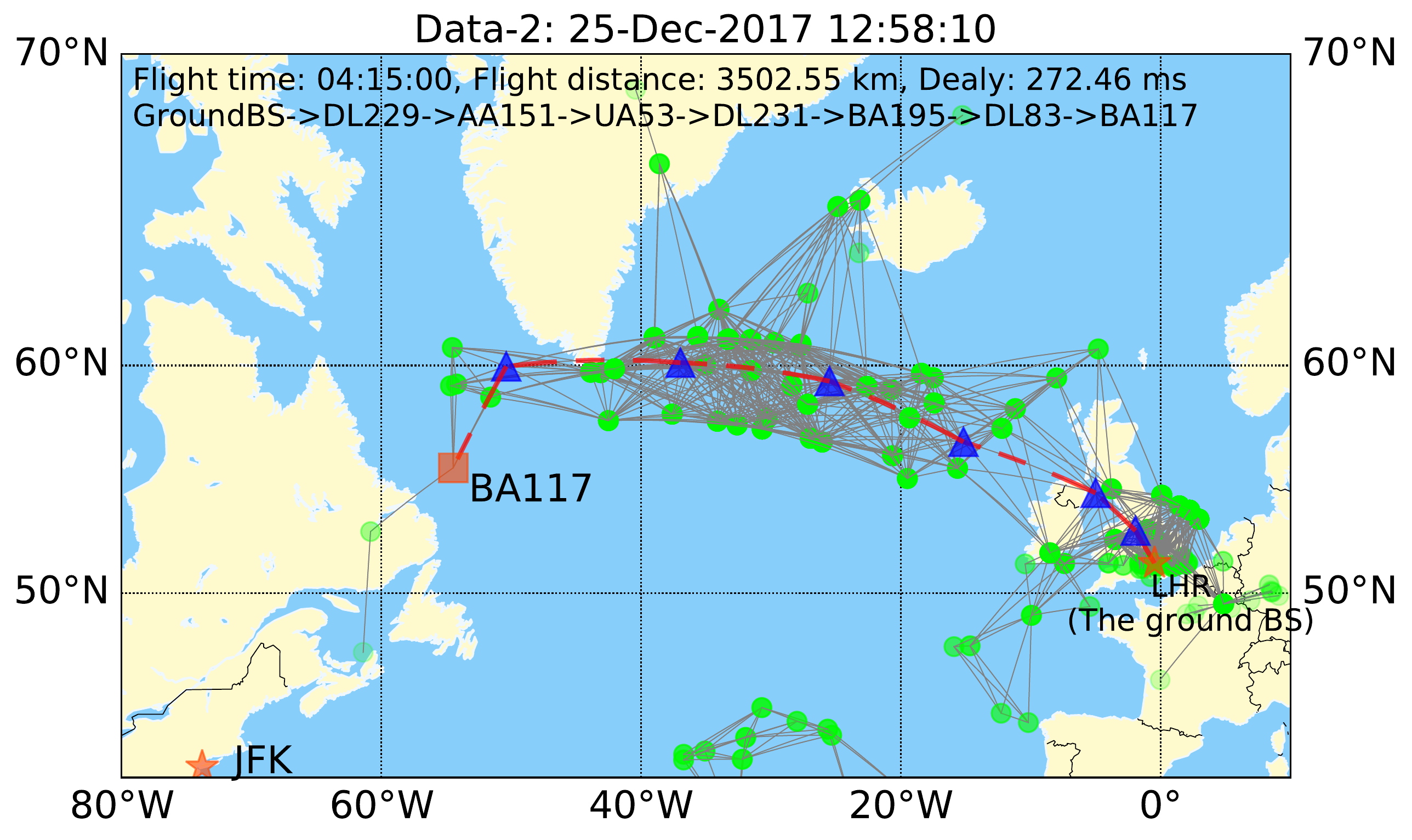}
 \label{fig:mapsub1}}
\subfigure[Results on Data-3]{
\includegraphics[ width = 0.45\linewidth]{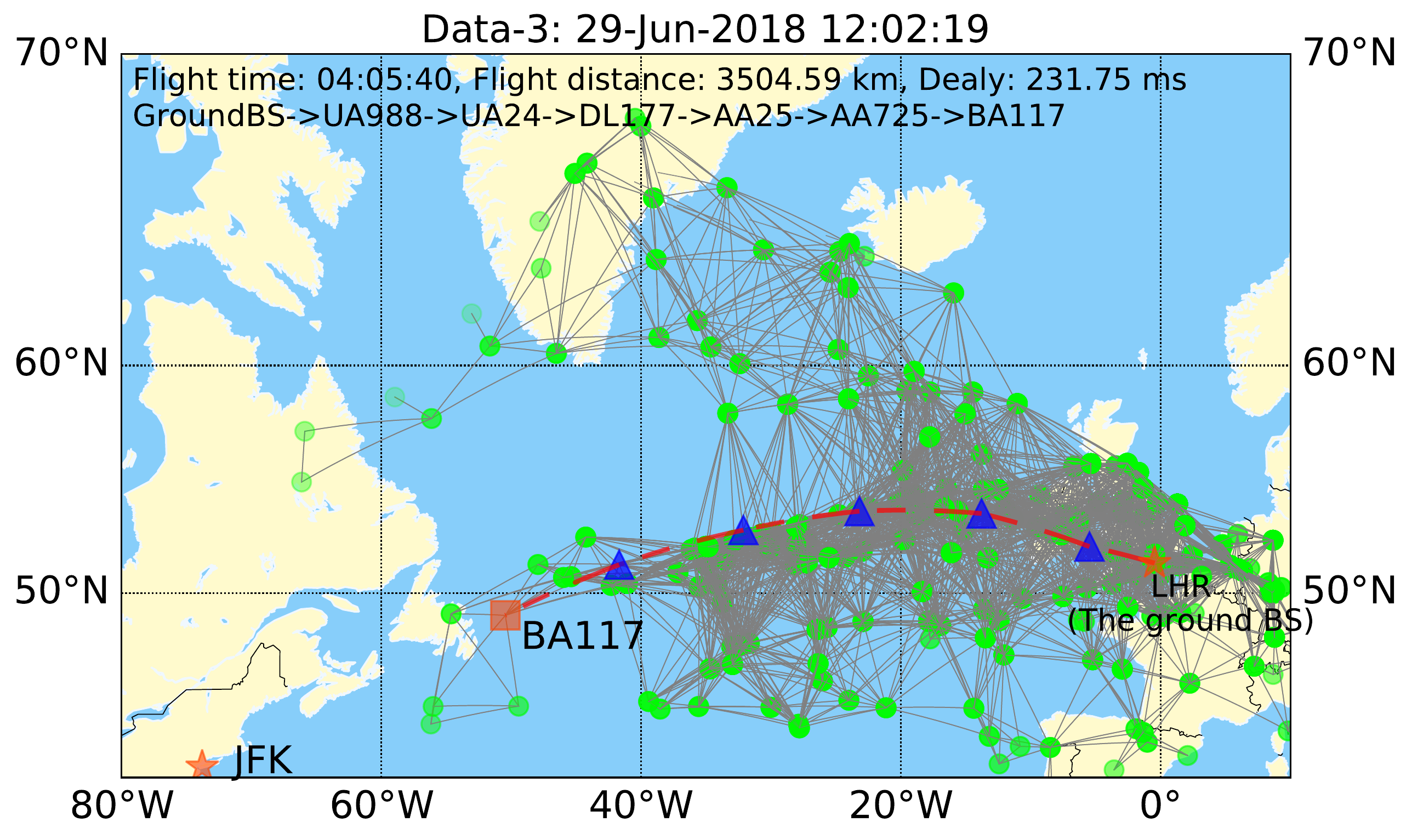}
 \label{fig:mapsub2}}
  \vspace{-0.5em}  \caption{\small Comparisons of BA117 in AANETs over different datasets.}
   \vspace{-1.0em}
   \label{fig:mapdata}
\end{figure}

Fig. \ref{fig:mapdata} shows an example topology of the AANET attained from Data-2 and Data-3 as well as the  shortest paths found  from the ground BS at LHR to BA117, when the flight distance of BA117 is  3532 km and each link has a fixed file-transfer rate $C = 10$Mbps. In Fig. \ref{fig:mapdata}, the stars denote LHR (the ground BS) and JFK, while green circles denote the planes in the network. Furthermore,  the red dashed lines denote the shortest path to BA117 (the square) and the triangles are the intermediate planes.   As seen from Fig. \ref{fig:mapsub1}, the shortest path from the ground BS at LHR to BA117 in Data-2 has seven hops (Ground BS$\rightarrow$DL229$\rightarrow$AA151$\rightarrow$UA53$\rightarrow$DL231$\rightarrow$BA195$\rightarrow$ DL83$\rightarrow$BA117) and its delay is $272.46$ms,  while  the shortest path found for BA117 in Data-3  has six hops (GroundBS$\rightarrow$UA988$\rightarrow$UA24$\rightarrow$DL177$\rightarrow$AA25$\rightarrow$AA725$\rightarrow$BA117) and its delay is $231.75$ms in Fig. \ref{fig:mapsub2}.  We can see that both the delay and the hop count of the shortest path found for Data-3 is lower than that for Data-2, which confirms that the AANET has  lower delay   during the busiest day than that in the quietest day of the year.  
This is owing to the fact that there are more flights in Data-3 than  in Data-2, since having more planes available  approaches the ideal scenario of Scheme 1.
 
\begin{figure} [t!]
\centering
\includegraphics[trim={0.3cm 0.1cm 0.1cm 0.1cm},clip, width = 0.45\linewidth]{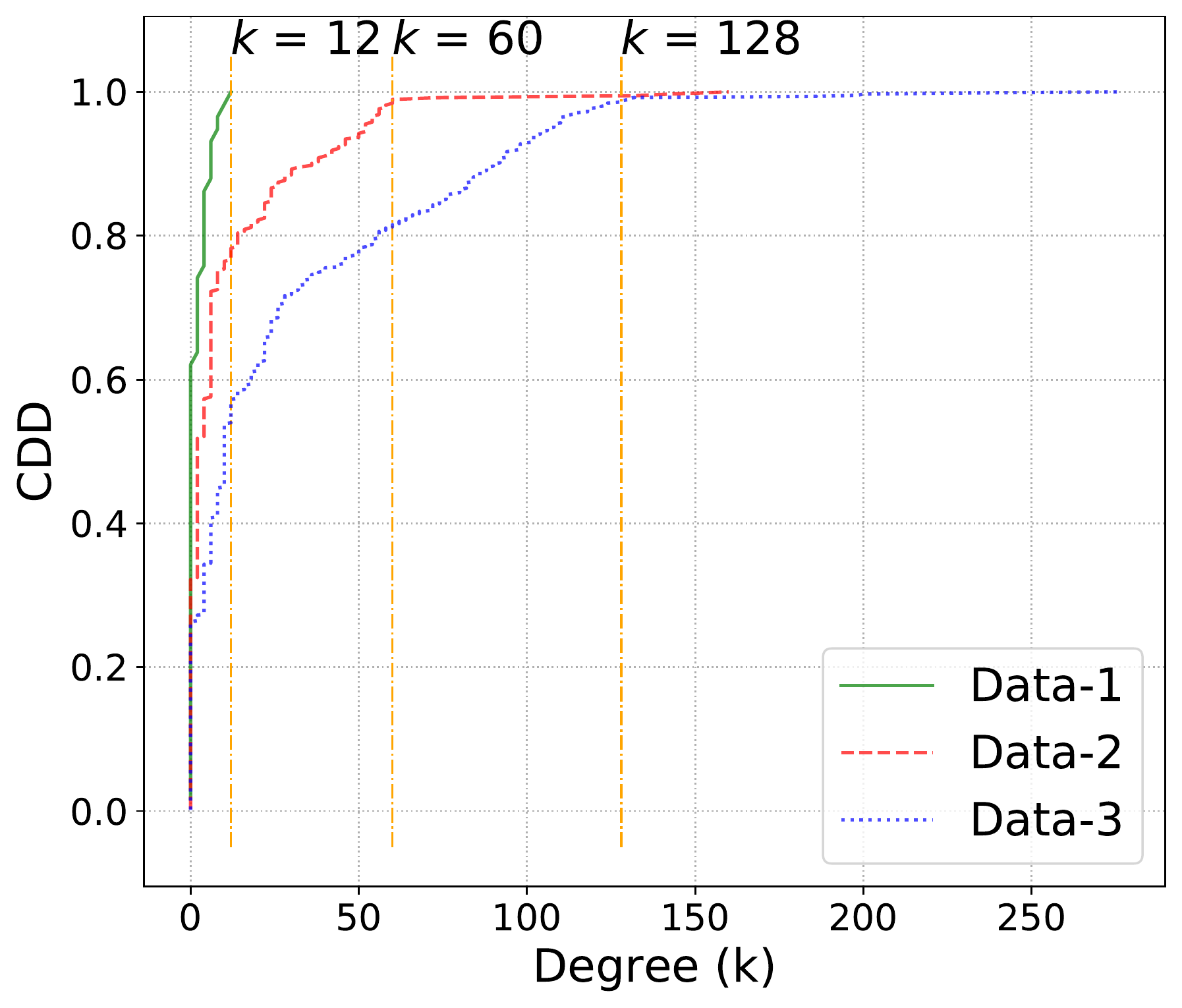}
  \vspace{-0.5em}  \caption{\small Degree distribution of AANET  with different datasets.}
  \label{fig:degree}
   \vspace{-1.0em}
\end{figure}

 Fig. \ref{fig:degree} shows the cumulative degree distribution (CDD) of the connected graph modelling  different networks generated from the three datasets for providing us with a glimpse into the structure of a network, where the  CDD represents the specific  fraction of nodes having a  degree smaller than a given number $k$ on the abscissa axis.  We observe that there are many nodes with degree zero for Data-1, which implies that there is no communication link between the ground BS at LHR  and BA117.  Furthermore, we can see that the probability of the nodes with  the degree of $k \ge 12$  is about 20\% for Data-2, while  it  is higher than 40\% for Data-3.  Moreover,  the maximum degree  of a node in Fig. \ref{fig:degree}  for Data-3 is 276, while it is 160 for Data-2 and 12 for Data-1, respectively.

\begin{figure}[!tp]
\centering
 \subfigure[CDFs of hops in  the shortest paths ]{
\includegraphics[ width = 0.45\linewidth]{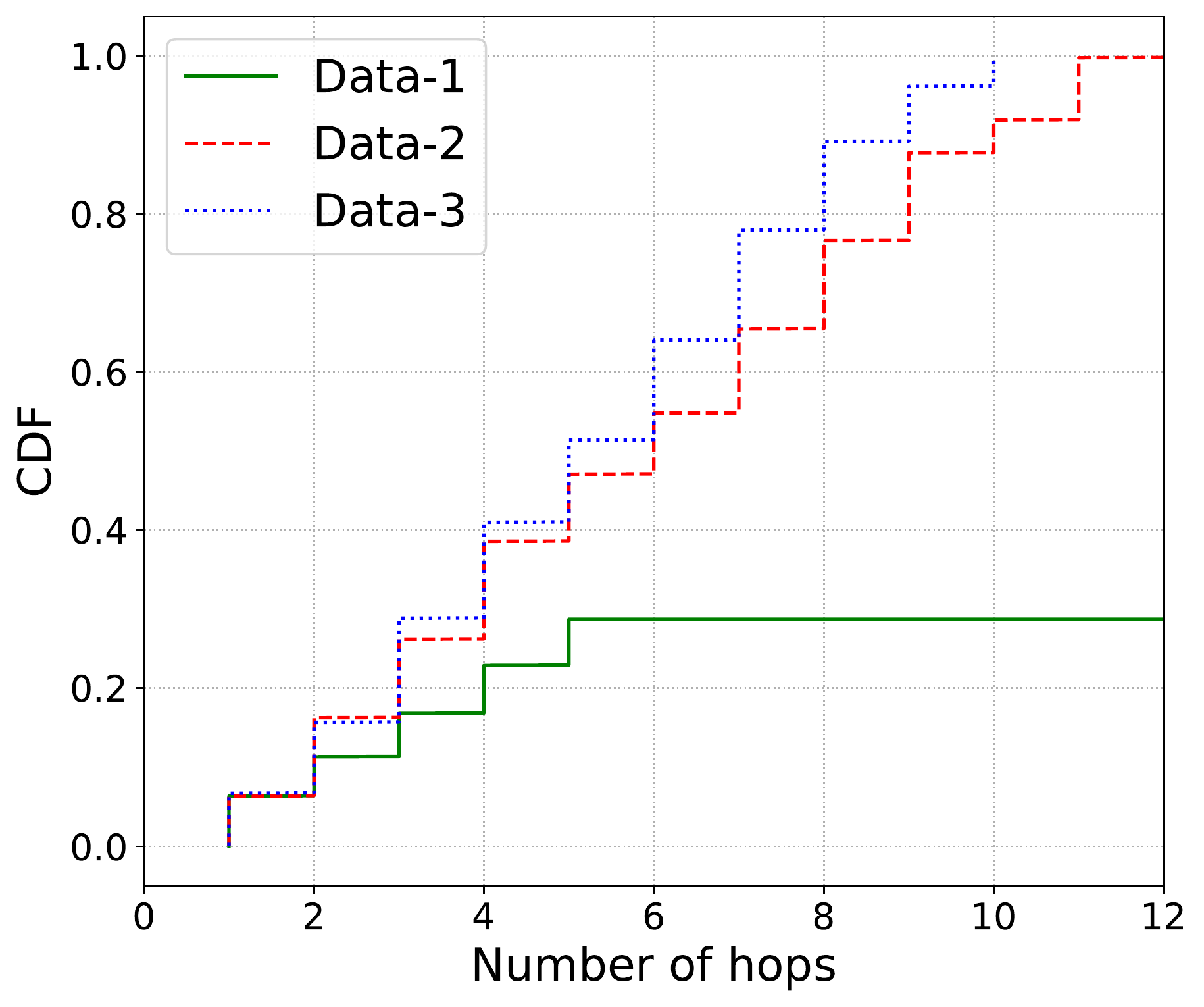}
 \label{fig:hopscdf}}
  \subfigure[CDFs of the delay in  the shortest paths]{
\includegraphics[width = 0.45\linewidth]{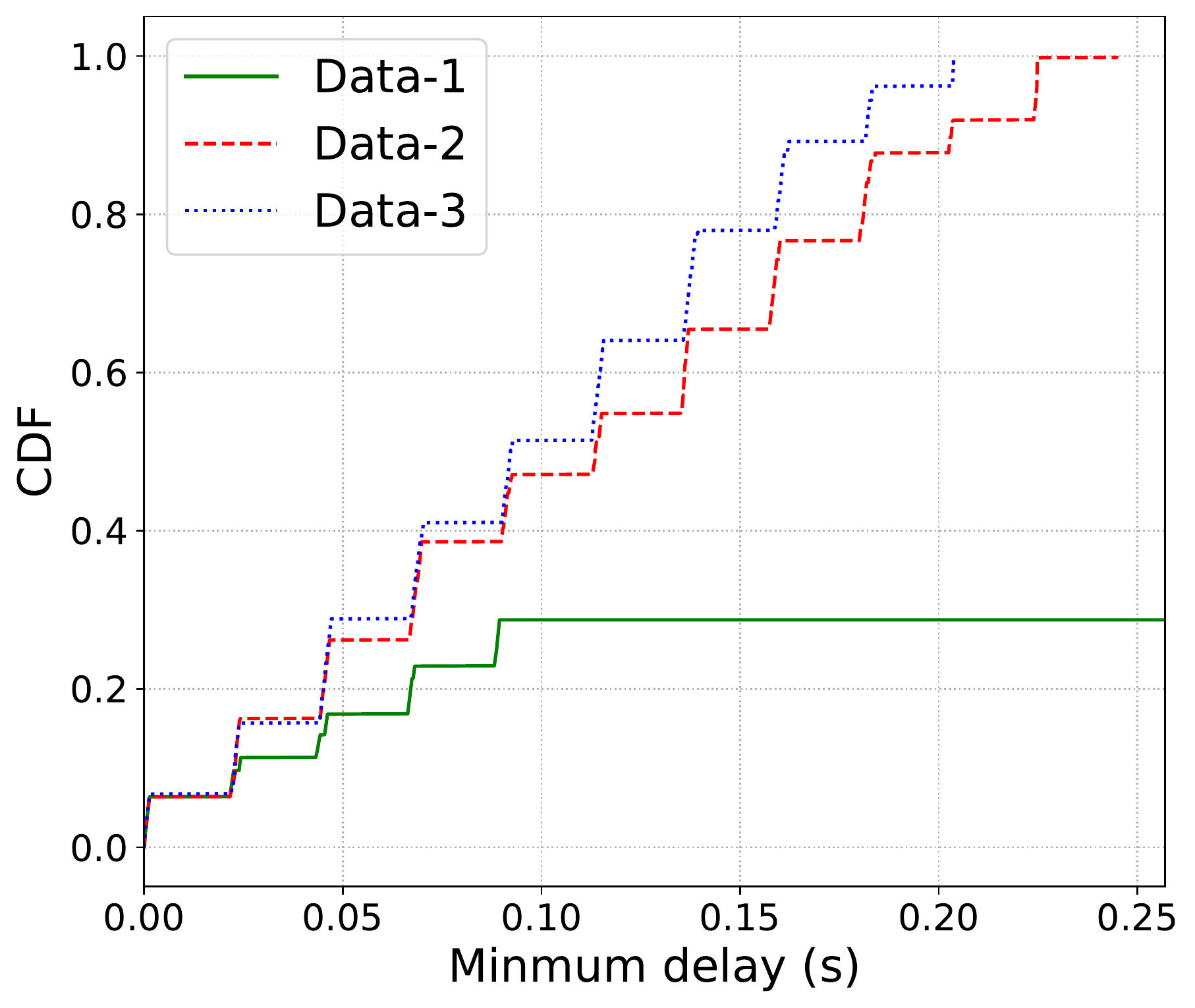}
 \label{fig:delaycdf}}
 \vspace{-0.5em}    \caption{ \small  CDFs of hops and the total delay of the shortest paths found by the proposed algorithm.}
   \vspace{-1.0em}
   \label{fig:linescom1}
\end{figure}

Fig. \ref{fig:linescom1} shows the  cumulative distribution functions (CDFs) of both the hop count  and of the total delay of  the shortest paths found  during the period of the complete travel of BA117 from LHR to JFK. 
Explicitly, Fig. \ref{fig:hopscdf} and Fig. \ref{fig:delaycdf} shows the cumulative probability distributions of the hop count and the total delay  in the shortest paths found in the three datasets throughout the complete travel of BA117.  
It can be observed  from the delay model of \eqref{eq:delaytot} that   the total  delay of a route relies on the number of hops.  As a result,   the CDF curves of the hop count have the similar trends as those of the total delay in terms of the shortest paths found.  
Furthermore,  Fig. \ref{fig:linescom1} also illustrates that  both the hop count and the delay attained from Data-3 are more beneficial than those of  Data-2 and Data-1, since there are more available flights in Data-3. 
More specifically,  observe from Fig. \ref{fig:linescom1} that during the complete period of travel there is a low success probability of around  30\%  for  BA117 to communicate with the ground BS via AANETs for Data-1. By contrast,  for Data-2 and Data-3,  BA117 is capable of connecting to the ground BS via multi-hop communications during its complete travel.

\vspace{-0.7em}
\section{Conclusions}
In this paper, we investigated the routing problem of AANET-assisted  integrated ground-air-space   communications with the objective of  providing in-flight connectivity. Whilst relying  both on aircraft and ground BSs as well as satellites, we minimized the total delay  by considering the propagation delay, file-transfer delay  and DF relaying delay. 
 Furthermore,  we have developed a weighted digraph for modelling the integrated AANET under minimum-rate constraints for finding the optimal route.
  Both the simulation results and real historical flight data driven results revealed that the  AANET-aided transmission has the potential benefit of extending the DA2GC coverage while providing reduced-delay transmissions. A promising extension of this research is to consider the multi-component Pareto-optimization of  AANET-assisted integrated networks for striking a compelling tradeoff amongst different performance metrics such as the delay, the energy efficiency as well as spectral efficiency, just to name a few.

\vspace{-0.9em}
{\small
\bibliographystyle{IEEEtran}
  \linespread{1.1}\selectfont
%PC use this path
%\bibliography{C:/Users/jc12n19/Dropbox/EndnoteLib/myrefv1}
%laptop use this path
\bibliography{mybib}
}
\end{document}